\documentclass[conference]{IEEEtran}
\IEEEoverridecommandlockouts
\usepackage{cite}
\usepackage{amsmath,amssymb,amsfonts}
\usepackage{algorithmic}
\usepackage{graphicx}
\usepackage{textcomp}
\usepackage{xcolor}
\usepackage{array}
\usepackage{float}
\usepackage{booktabs}
\usepackage{multirow}
\usepackage{booktabs,subcaption}

\newcommand{\bc}{\textcolor[rgb]{0,0,0}}
\newcommand{\rc}{\textcolor[rgb]{0,0,0}}
\newcommand{\gc}{\textcolor[rgb]{0,0,0}}

\def\BibTeX{{\rm B\kern-.05em{\sc i\kern-.025em b}\kern-.08em
    T\kern-.1667em\lower.7ex\hbox{E}\kern-.125emX}}
\begin{document}

\title{Leveraging Transfer Learning and User-Specific Updates for Rapid Training of BCI Decoders}

\author{Ziheng Chen$^{1,\star}$, Po T. Wang$^{1}$, Mina Ibrahim$^{1}$, Shivali Baveja$^{2}$, Rong Mu$^{3}$, An H. Do$^{4}$, and Zoran Nenadic$^{1,3,\star}$
\thanks{$^{1}$Department of Biomedical Engineering, University of California, Irvine (UCI), CA, USA $^{\star}$\tt\small zihengc5@uci.edu, znenadic@uci.edu}
\thanks{$^{2}$School of Medicine, UCI}
\thanks{$^{3}$Department of Electrical Engineering and Computer Science, UCI}
\thanks{$^{4}$Department of Neurology, UCI}
}
\maketitle
\begin{abstract}
Lengthy subject- or session-specific data acquisition and calibration remain a key barrier to deploying electroencephalography (EEG)-based brain-computer interfaces (BCIs) outside the laboratory. Previous work has shown that cross-subject, cross-session invariant features exist in EEG. We propose a \emph{transfer learning pipeline} based on a two-layer convolutional neural network (CNN) that leverages these invariants to reduce the burden of data acquisition and calibration. A baseline model is trained on EEG data from five able-bodied individuals and then rapidly updated with a small amount of data from a sixth, holdout subject. The remaining holdout data were used to test the performance of both the baseline and updated models. We repeated this procedure via a leave-one-subject-out (LOSO) validation framework.
Averaged over six LOSO folds, the updated model improved classification accuracy upon the baseline by 10.0, 18.8, and 22.1 percentage points on two binary and one ternary classification tasks, respectively. These results demonstrate that decoding accuracy can be substantially improved with minimal subject-specific data. They also indicate that a CNN-based decoder can be personalized rapidly, enabling near plug-and-play BCI functionality for neurorehabilitation and other time-critical EEG applications.
\end{abstract}


\section{Introduction}\label{sec:intro} 
Electroencephalography (EEG)-based brain–computer interfaces (BCIs) have been increasingly used in neuroprosthetic or neurorehabilitation applications~\cite{pfurtscheller2003thought, do2013brain, king2015feasibility, ramos2013brain, mccrimmon2015brain, donati2016long}. However, long-term BCI deployment typically relies on repeated collection of training EEG data across time and participants. While this approach optimizes BCI operation across subjects and experimental sessions, its practical deployment is hindered by the time lost to collect data and retrain/recalibrate the BCI system~\cite{krauledat2006reducing,lotte2015signal, lotte2018review}. This is especially problematic in people with neurological injury, whose endurance and availability may be limited~\cite{jayaram2016transfer,lotte2018review}. The problem is further compounded in BCI clinical trials~\cite{biswas2024single}, where BCI retraining/recalibration detracts from valuable intervention time.

Besides time inefficiency, the above approach fails to take advantage of EEG features that may be common across subjects and experiments~\cite{krauledat2006reducing,jayaram2016transfer}. To utilize these invariant EEG features and train a general BCI decoder,  various transfer learning techniques have been proposed~\cite{lotte2018review}. However, these general models typically result in a lower decoding accuracy than subject- and session-specific decoders~\cite{krauledat2006reducing, fazli2009subject}. Therefore, transfer learning approaches generally lead to practicality and performance tradeoffs.

Motivated by this shortcoming, our study seeks to augment the zero-training transfer learning framework~\cite{krauledat2008towards} by updating the general BCI decoder with a small amount of subject-specific EEG data. Specifically, we trained a lightweight two-layer convolutional neural network (CNN) on multi-subject spatio-spectral EEG data. We then demonstrated through leave-one-subject-out (LOSO) validation that the performance of this general model can be significantly improved when the model is fine-tuned with a small amount (as little as 2\% of the original training sample) of the target subject's data. Moreover, unlike traditional machine learning approaches~\cite{jayaram2016transfer}, updating pre-trained neural network models can be done seamlessly, further highlighting the implementation efficiency of the proposed approach.

To demonstrate the efficacy of our framework, we collected EEG data from multiple subjects as they engaged in the occipital $\alpha$-wave modulation and event-related synchronization/desynchronization produced by an overt motor action. We then tested the performance of our models on binary and ternary classification tasks. Our findings suggest the performance gap between the general and subject-specific decoders can be closed, while eliminating the need for \textit{de novo} model training. Therefore, the proposed approach may lead to the rapid deployment of personalized EEG decoders with both clinical and non-clinical implications.

\section{Methods}\label{sec:methods}
\subsection{Overview}
\bc{Subjects were recruited and their EEG data were collected across two experiments. Subsequently, their data were organized in two binary and one ternary classification tasks. For each classification task, a CNN model was trained on all but one holdout subject's data. This model was then fine-tuned with 10\% of the holdout subject's data. Subsequently, both the basic and updated models were validated on the remaining holdout data (90\%). The above procedure was repeated within the LOSO validation framework, and the average performances of the basic and updated models are reported and compared.}

%-------------------------------------------------------------
\subsection{Subject Recruitment}
We obtained the approval from the Institutional Review Board at the University of California,
Irvine and solicited able-bodied individuals to participate in this study.

%-------------------------------------------------------------
\subsection{Data Collection}
We used a 32-channel EEG cap (Waveguard, ANT North America, Philadelphia, PA, USA) with International 10-20 Standard montage. We applied a conductive gel and skin abrasion to the following 15-electrode subset: Fp1, Fp2, F3, Fz, F4, C3, Cz, C4, P3, Pz, P4, POz, O1, Oz, and O2, with AFz as the reference. The 30-Hz contact impedances were kept below 5 k$\Omega$. Data were amplified ($\times$5,000), band-pass filtered (1-35 Hz), and digitized (200~Hz, 16 bits) using an EEG acquisition system (BIOPAC Systems Inc., Goleta, CA, USA).

Subjects participated in the following experimental tasks, while being guided by computer-controlled audio cues:
\begin{enumerate}
  \item \textbf{Experiment 1: Eyes-Open (EO) vs. Eyes-Closed (EC)}. Subjects alternated between keeping their eyes open and closed. The subjects were instructed to relax during EC periods, so that their occipital $\alpha$-wave could be established. Conversely, the $\alpha$-waves are expected to be disrupted during EO periods. 
  \item \textbf{Experiment 2: Idle vs. Fist Pump (FP).} Subjects alternated between rest periods and repetitive fist pumps, to elicit event-related synchronization/desynchronization over the contralateral motor cortex in the $\mu$ and $\beta$ physiological bands~\cite{king2014performance}.
\end{enumerate}

Each experiment consisted of four runs with short ($\sim$5 min) breaks in between. Each run generated 500 s of EEG data, where subjects alternated between the EO and EC states or Idle and FP states every 10 s. Subjects followed auditory cues and were monitored by an operator for experimental compliance.

%-------------------------------------------------------------
\subsection{Data Pre-processing}

For each 10-s-long trial, the initial and final second of data were removed to account for subject reaction time and signal transients. The remaining 8 s of data were further segmented into two non-overlapping 4-s-long windows, resulting in $15\times800$ time-domain matrices (15 channels, 800 time samples). Each data window was then represented in the frequency domain using the Fast Fourier Transform (frequency resolution 0.25 Hz), and the magnitude spectra were calculated over the non-negative frequency bands, resulting in $15\times400$ frequency-domain feature matrices.

From these data and their corresponding behavioral labels, as determined by audio cues, we constructed the following classification tasks (see Table~\ref{tab:tasks}). The classifications Tasks 1 and 2, respectively, matched the mental states observed in Experiments 1 and 2. To test the performance of our method beyond binary classification, we combined data from Experiments 1 and 2 and formed a ternary classification task (Task 3). To this end, we combined the EO and Idle states into a single class due to their behavioral similarity, while keeping the EC and FP states as separate classes.

\begin{table}[!htbp]
\caption{The definition of classification tasks.}
\label{tab:tasks}
\centering
\begin{tabular}{lll}
\toprule
\textbf{Task} & \textbf{Data Source} & \textbf{Classes}\\
\midrule
1	&	Experiment 1 &	class 1 = \{EO\}\\ 
 & & class 2 = \{EC\}\\
 \addlinespace
2	&	 Experiment 2	&	 class 1 = \{Idle\}\\
& & class 2 = \{FP\}\\
\addlinespace
3 &	 Experiment 1	&	  class 1 = \{EO, Idle\} \\
 & Experiment 2 & 
class 2 = \{EC\}\\
& & class 3 = \{FP\}\\
\bottomrule
\end{tabular}
\end{table}

\subsection{CNN Model Architecture}
\begin{figure}[!htbp]
  \centering
\includegraphics[width=\linewidth]{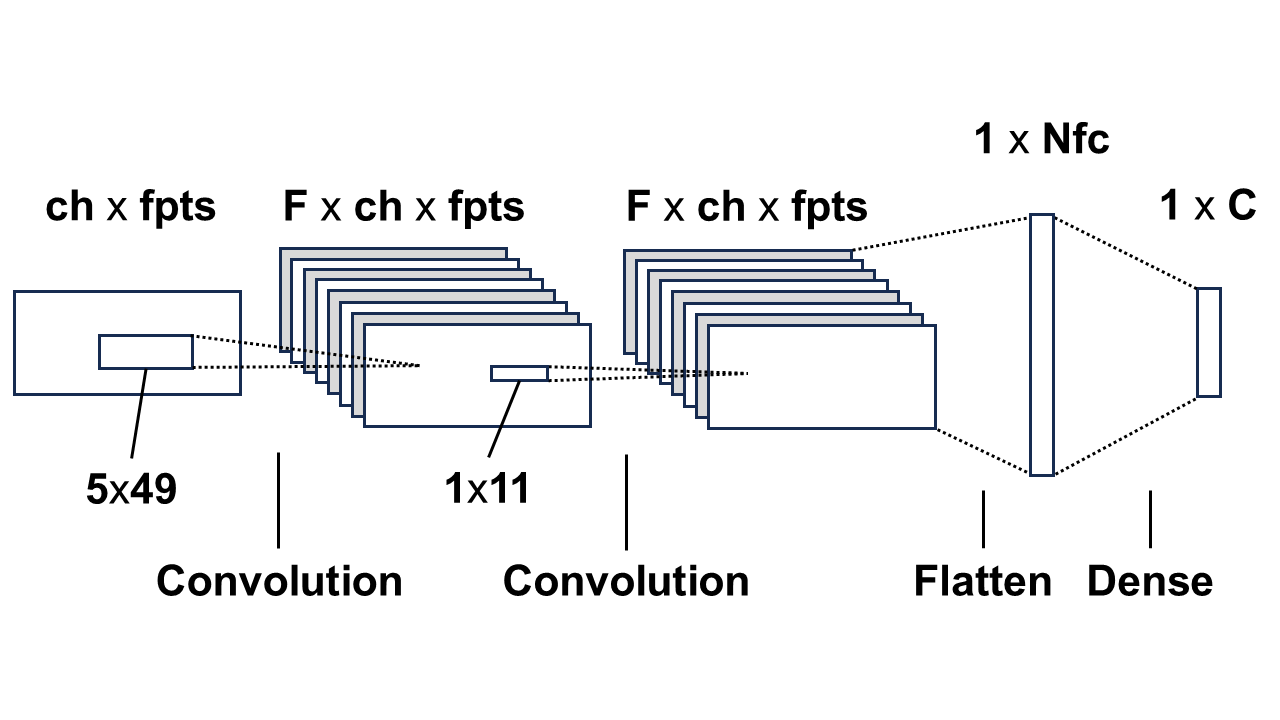}
  \caption{Two-layer CNN architecture. The input is a spatio-spectral feature matrix  ($\text{ch}\times\text{fpts}$), where
  $\text{ch}$ is the number of EEG channels and  $\text{fpts}$ is the number of frequency points.
    The first 2D convolutional layer applies $\text{F}$ filters (filter size $5\times49$) with rectified linear unit (ReLU) activation function to extract spatio-spectral features, producing an $\text{F}\times\text{ch} \times \text{fpts}$ output. The second convolutional layer uses $1\times 11$ filters with ReLU activation to model within-channel spectral dynamics. Subsequently, a dropout layer ($p = 0.5$) is applied, the features are flattened, and passed to a fully-connected layer with $\text{Nfc}$ ReLU-activated units. Finally, a $1\times \text{C}$ output layer is added for classification,  where $\text{C}$ is the number of classes. The network parameters are given generally. In the present study $\text{ch}=15, \text{fpts}=400, \text{F}=8, \text{Nfc}=128$, and $\text{C}=2$ for binary cases and $\text{C}=3$ for ternary case.}
  \label{fig:cnn_architecture}
\end{figure}

\bc{Figure~\ref{fig:cnn_architecture} summarizes the network's model architecture. A two-layer CNN employing 2D convolutional layers (Adam optimizer, cross entropy loss, batch size 64) was implemented using PyTorch~\cite{paszke2019pytorch}. Its inputs are $15\times400$ spatio-spectral EEG features. The first convolutional layer applied eight $5 \times 49$ filters, thereby combining data from five EEG channels and $\sim$12 Hz frequency bands. 
This matches the width of common physiological sub-bands, especially in the $\beta$ band.
These operations were followed by a nonlinear activation function from the rectified linear unit (ReLU) family. A second convolutional layer with eight $1\times11$ filters further refined the spectral details without additional spatial mixing. This operation was also followed by the ReLU activation function. Subsequently, we applied a dropout rate of $p = 0.5$ to mitigate over-fitting~\cite{srivastava2014dropout}. The resulting feature map was then flattened and forwarded to a fully connected layer of 128 ReLU-activated units, culminating in an output layer with two neurons for binary tasks (Task 1 and Task 2) or three neurons for Task 3.} All numerical design choices, including filter (kernel) sizes, filter counts, learning rate, dropout probability and fully connected units, were selected through hyperparameter tuning on the training data.

\subsection{Model Training and Validation}

\begin{figure}[ht]
\centering
\includegraphics[width=\linewidth]{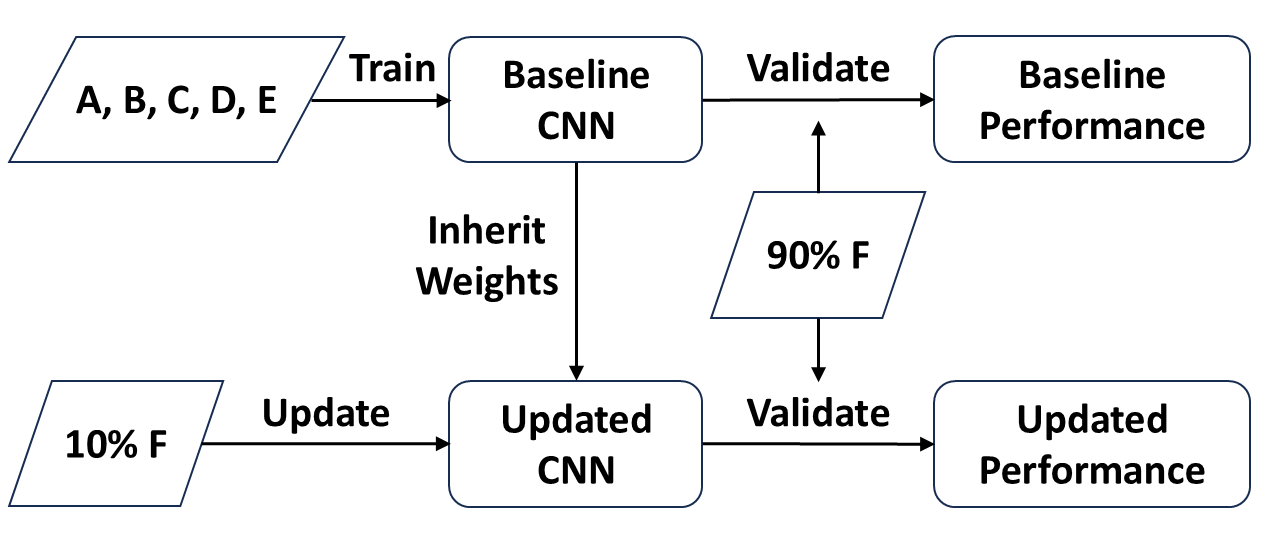}
  \caption{Model training and validation pipeline. EEG data from subjects A-E were used to train the baseline model, which learned generalizable spatio-spectral EEG features for a given classification task. Subject F is held out for updating the baseline model and evaluation. We initialized the updated model with the parameters of the pre-trained baseline model and then resumed the training process by providing 10\% of subject F's data. Both the original baseline model and the updated personalized model were then evaluated on the remaining 90\% of subject F's data. This procedure was repeated in a LOSO setup, each time designating a different subject for a holdout.}
  \label{fig:cnn_pipeline}
\end{figure}

Model training and validation were carried out in a LOSO loop repeated across all subjects, as shown in Figure~\ref{fig:cnn_pipeline}. Briefly, EEG data from subjects A-E were used to train the baseline model to learn generalizable spatio-spectral features for a given classification task. One subject (subject F) was designated as a holdout, and 10\% of their data were selected via stratified shuffle split and used to update the baseline model. Before fine-tuning, the updated model was initialized with the parameters of the pre-trained baseline model. The remaining 90\% of subject F's data were used to evaluate the performance of the baseline and updated models. We used class-weighted cross-entropy loss to mitigate potential residual class imbalances~\cite{torchCrossEntropy}. The above procedures were repeated in a LOSO setup, each time designating a different subject for a holdout.

We employed an adaptive stopping criterion to terminate training when additional training epochs failed to yield meaningful loss function reduction. Specifically, we halted training when the average loss over a sliding epoch window failed to improve by a specified relative margin for a consecutive span defined by the patience parameter. Following an initial warm-up period, the algorithm tracked windowed loss averages and maintained a running best. If no sufficient improvements were observed, and no recent spikes in loss function were detected, a wait counter was incremented. Training stopped once this counter exceeded the patience threshold and the latest loss did not exceed the previous training epoch’s loss, indicating diminishing returns from continued training.

To ensure reproducibility, deterministic training procedures were employed by fixing random seeds across all random modules, and by configuring PyTorch to use deterministic GPU operations only~\cite{pytorch_reproducibility}.

Model performance was quantified using overall classification accuracy for all three tasks. Sensitivity and specificity were used for Task 1 and Task 2 to capture true positive and true negative rates (TPR and TNR), respectively. For the three-class task, row-normalized confusion tables where the diagonal represents the sensitivity were calculated. All metrics were evaluated on 90\% of the holdout data designated for testing and aggregated across all holdout subjects. Since we propose to update the baseline model with subject-specific data, it is worth investigating how our approach compares to a model solely based on subject-specific data. To this end, a standard SVM algorithm with a radial basis function kernel was trained on the same 10\% of the holdout data and tested on the remaining 90\%. This procedure was then repeated for all subjects. This approach simulates a common BCI practice of \textit{de novo} model training across subjects and experiments. We chose the SVM classifier since 10\% of the holdout data was not sufficient to train the CNN network with the architecture shown in Fig.~\ref{fig:cnn_architecture}.

\section{Results}\label{sec:results}

\begin{figure}[ht]
  \centering
\includegraphics[width=\linewidth]{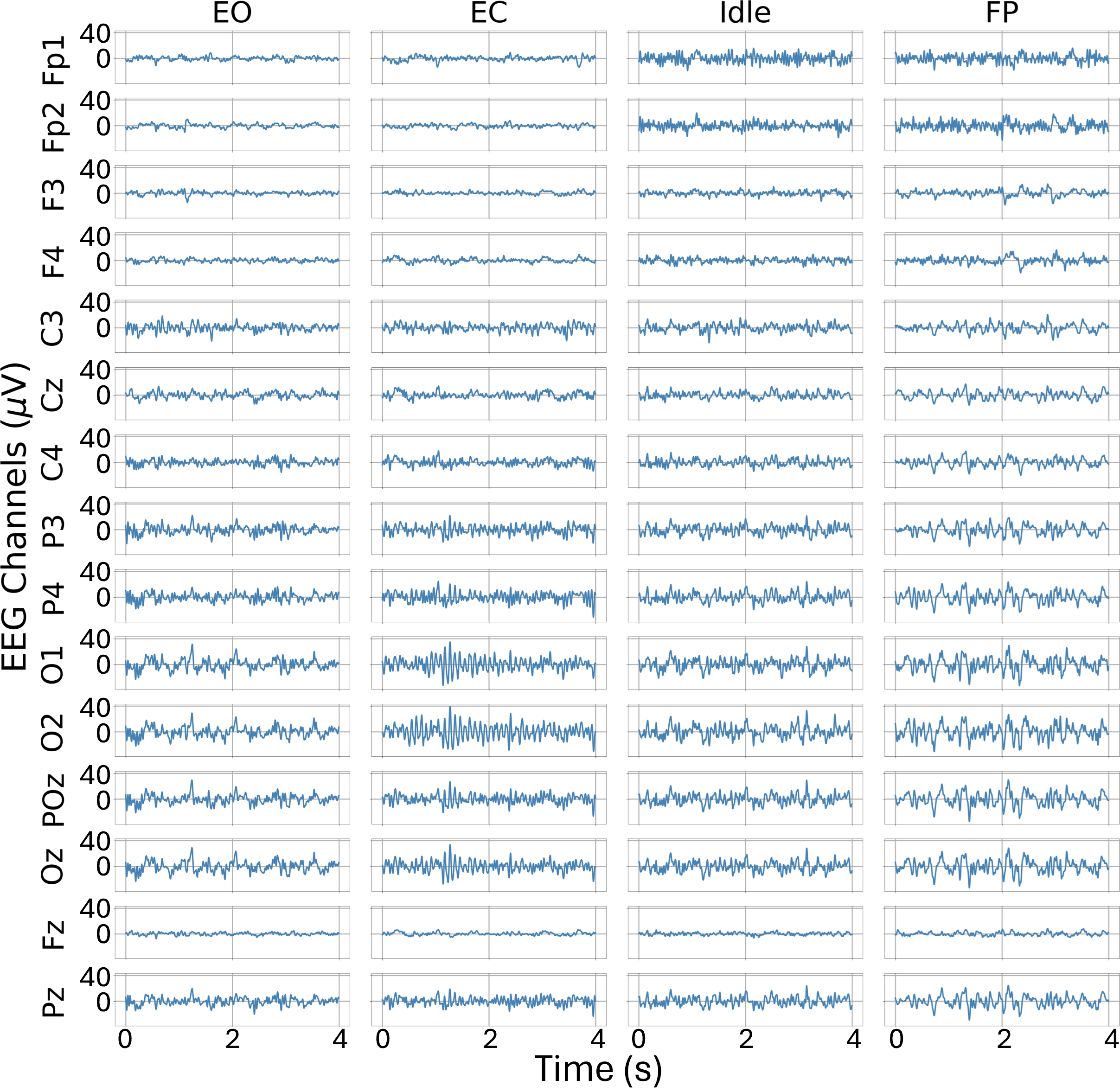}
  \caption{Representative EEG samples from each experimental task. 15 EEG channels (Fp1, Fp2, F3, Fz, F4, C3, Cz, C4, P3, Pz, P4, POz, O1, Oz, and O2) from Subject 1 (Experiment 1) and Subject 2 (Experiment 2) are shown.}
  \label{fig:ex_eeg}
\end{figure}

Six able-bodied participants (3F, 3M, age 29$\pm$11) were recruited and signed an informed consent to participate in the study. The subjects could follow instructions with a high degree of compliance and completed the two experimental protocols within a single day. Their EEG data were labeled by the auditory cue and stored for further offline analysis. Representative examples are shown in Fig.~\ref{fig:ex_eeg}. The data were subsequently pre-processed and analyzed according to the pipeline described in Section~\ref{sec:methods}.

\subsection{Training Dynamics}
Model training followed a LOSO transfer learning protocol. Briefly, the data from five subjects were used to train the baseline CNN model, with the sixth subject's data held out for update and validation (Fig.~\ref{fig:cnn_pipeline}). Specifically, we selected 10\% of the holdout data via stratified sampling to preserve class balance. These data were used to fine-tune (personalize) the baseline models. We then evaluated the performance of baseline and personalized models on the remaining 90\% of the holdout data. Class-weighted cross-entropy loss was employed during both the baseline and personalized training phases to address residual class imbalance, especially in Task 3. The above procedure was repeated six times, each time selecting a different subject as a holdout.

\begin{figure}[!htbp]
  \centering
\includegraphics[width=\linewidth]{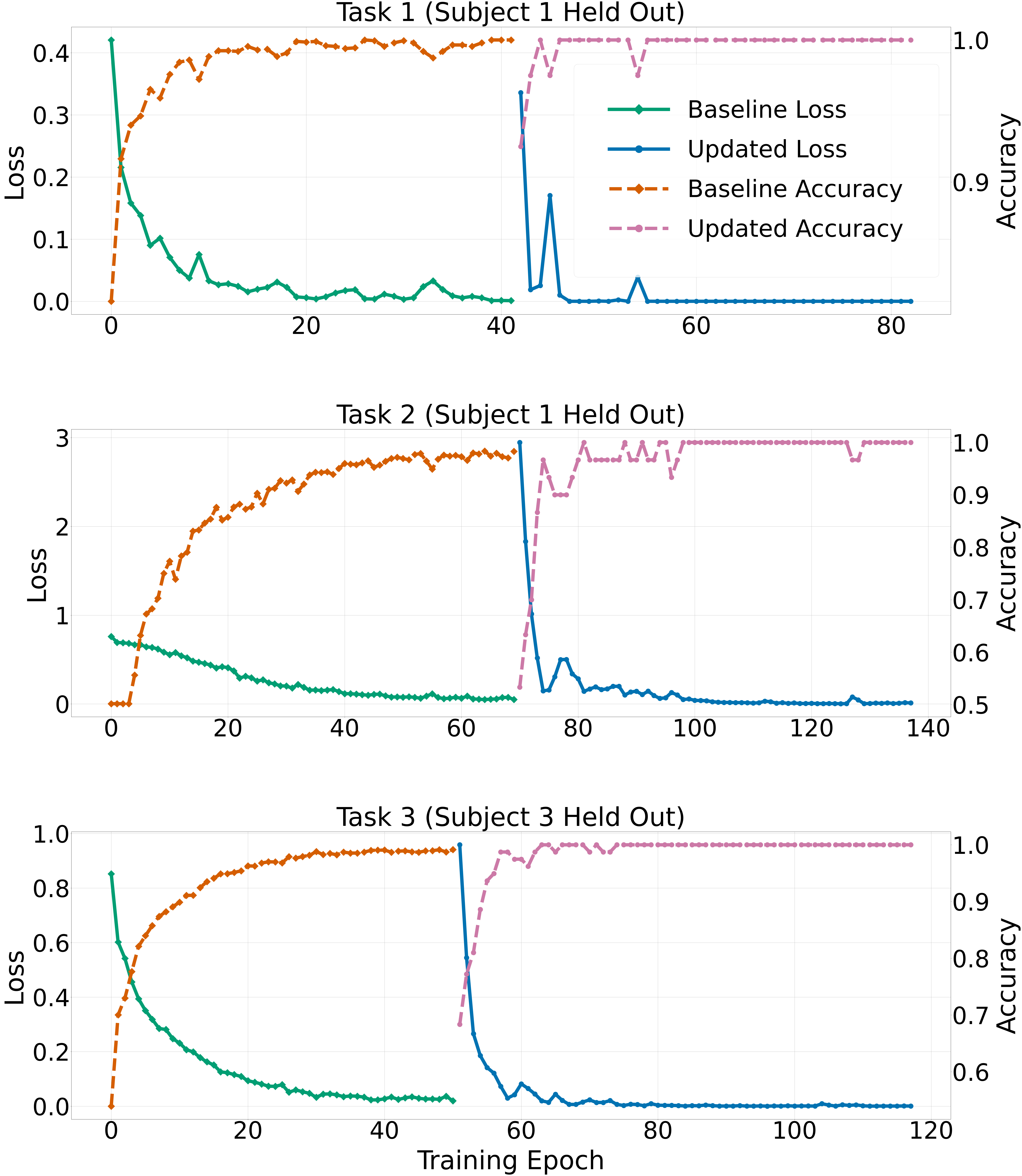}
  \caption{Representative plots of the loss function and accuracy over training epochs for both the baseline and updated models and across the three classification tasks. \textbf{Top:} Task~1 (EO vs.\ EC) showing loss and accuracy trajectories for both the baseline model (trained without Subject 1's data) and the personalized model (fine-tuned with 10\% of Subject 1's data). \textbf{Middle:} Similar convergence patterns for Task 2 (Idle vs.\ FP) classification model, with Subject 1's data held out for update. \textbf{Bottom:} Equivalent plot for Task~3 (ternary classification) model with Subject 3's data held out for update.}
  \label{fig:convergence_ex}
\end{figure}

Training loss function and accuracy were monitored across training epochs. For each CNN model, the stopping criteria terminated training before reaching the maximum training epoch limit.  Figure~\ref{fig:convergence_ex} shows representative examples of these functions for each of the three classification tasks. Generally, the loss function decreased with training, albeit with different rates and terminating at different numbers of training epochs. Once the update data were introduced, the loss function momentarily increased, before rapidly decaying and eventually reaching the stopping criterion.

\subsection{Classification Accuracy}
Updated models consistently outperformed baseline models across all three classification problems (Table~\ref{tab:accuracy}). For Tasks 1, 2, and 3, the performance gains were 10.0, 18.8, and 22.1 percentage points, respectively. Paired two-tailed t-tests confirmed that these gains were statistically significant for each task: EO vs.\ EC ($p = 0.047$), Idle vs.\ FP ($p = 0.011$), and the three-class task ($p = 0.002$). 
 These results highlight the effectiveness of subject-specific adaptation in enhancing decoding performance for both binary and ternary EEG classification tasks. Updated models also outperformed the subject-specific SVM models by a significant margin, especially in Tasks 2 and 3 (see Table~\ref{tab:svm_accuracy}). Thus, the subject-specific data can generate a more informative model when placed in the context of a larger and more diverse dataset.

\begin{table}[!htbp]
\caption{CNN mean classification accuracy (\%) across six subjects, with each model's performance estimated through a LOSO procedure. Values are shown as mean~$\pm$~standard deviation, weighted by each holdout subject's sample size.}
\label{tab:accuracy}
\centering
\begin{tabular}{lll}
\toprule
\textbf{Task} & \textbf{Baseline Model} & \textbf{Updated Model}\\
\midrule
1 	&	 83.84$\pm$11.09	&	 93.81$\pm$5.77	\\
2	&	 63.47$\pm$\,\,\,6.08	&	 82.26$\pm$6.53	\\
3	&	 61.23$\pm$\,\,\,9.29	&	 83.30$\pm$5.63	\\
\bottomrule
\end{tabular}
\end{table}

\begin{table}[h]
\caption{\rc{SVM mean classification accuracy (\%) across six held-out subjects, trained on the 10\% hold-out dataset used to update the CNN models. Values are shown as mean~$\pm$~standard deviation, weighted by each holdout subject's sample size.}}
\label{tab:svm_accuracy}
\centering
\setlength{\tabcolsep}{12pt}
\begin{tabular}{lcc}
\toprule
\textbf{Task} & \textbf{SVM Model} \\
\midrule
1 	&	 89.43$\pm$5.71 \\ 
2	&	 60.40$\pm$4.68 \\
3	&	 72.25$\pm$6.07 \\
\bottomrule
\end{tabular}
\end{table}

\subsection{Per-Class Sensitivity and Specificity}

Substantial per-class improvements were observed following the update with the held out subject's data (see Table~\ref{tab:sensspec_task1}). For Task 1 (EO vs.\ EC), sensitivity and specificity increased by 10.02 and 9.90 percentage points, respectively, showing an improvement in differentiating the presence/absence of the occipital $\alpha$-wave. In Task 2 (Idle vs. FP), sensitivity and specificity improved by 17.37 and 20.22 percentage points, respectively, underscoring the decoder’s enhanced ability to differentiate overt motor actions from rest states. The higher sensitivity and specificity of the baseline model in Task 1 are consistent with the occipital $\alpha$-wave being more robustly present than event-related desynchronization underlying motor execution.

\begin{table}[h]
\caption{CNN per-class mean LOSO sensitivity and specificity (\%) across six subjects for Tasks 1 \& 2. EO and Idle are considered the default state.}
\label{tab:sensspec_task1}
\centering
\setlength{\tabcolsep}{4.5pt}
\begin{tabular}{lcccc}
\toprule
\textbf{Task} & \multicolumn{2}{c}{\textbf{Baseline Model}} & \multicolumn{2}{c}{\textbf{Updated Model}} \\
\cmidrule(lr){2-3} \cmidrule(lr){4-5}
 & Sens. & Spec. & Sens. & Spec. \\
\midrule
1	&	 82.64$\pm$14.66	&	 85.05$\pm$24.05	&	 92.66$\pm$7.07	&	 94.95$\pm$6.45	\\
\midrule
2	&	 69.60$\pm$30.56	&	 57.32$\pm$27.19	&	 86.97$\pm$3.74	&	 77.54$\pm$11.69	\\
\bottomrule
\end{tabular}
\end{table}

For Task 3, the most significant gains were observed in the FP class. Sensitivity increased 31.62 percentage points (see Table~\ref{tab:cnn_cm_task3}), highlighting the decoder’s improved ability to detect EEG signals underlying motor actions. EO/Idle sensitivity rose 25.19 percentage points, and EC sensitivity remained high.

\begin{table}[!htbp]
    \centering
    \caption{Row-normalized (\%) confusion table for baseline and updated CNN models across six subjects for Task 3. }
    \subcaption*{(a) \textbf{Baseline model}}
    \label{tab:cnn_cm_task3}
    \begin{tabular}{lccc}
        \toprule
        True Class & \multicolumn{3}{c}{Predicted Class}\\\cmidrule(lr){2-4}
        & EO, Idle & EC &  FP \\
        \midrule
EO, Idle	&	 60.20$\pm$23.38	&	 12.82$\pm$20.21	&	 26.97$\pm$24.12	\\
EC	&	  \,\,\,9.29$\pm$\,\,\,7.41	&	 82.90$\pm$10.78	&	  \,\,\,7.81$\pm$\,\,\,7.42	\\
FP	&	 50.11$\pm$23.89	&	 12.36$\pm$21.16	&	 37.53$\pm$22.44	\\
        \bottomrule
    \end{tabular}
    \vspace{1em}

    \subcaption*{(b) \textbf{Updated model}}
    \begin{tabular}{lccc}
        \toprule
        True Class & \multicolumn{3}{c}{Predicted Class}\\\cmidrule(lr){2-4}
        & EO, Idle & EC &  FP \\
        \midrule
EO, Idle	&	 85.39$\pm$\,\,\,7.06	&	  \,\,\,3.26$\pm$\,\,\,1.63	&	 11.35$\pm$\,\,\,5.78	\\
EC	&	  \,\,\,7.62$\pm$\,\,\,9.71	&	 91.26$\pm$10.68	&	  \,\,\,1.12$\pm$\,\,\,1.79	\\
FP	&	 29.84$\pm$12.16	&	  \,\,\,1.00$\pm$\,\,\,1.19	&	 69.15$\pm$12.34	\\
        \bottomrule
    \end{tabular}
\end{table}

\section{Discussion}
This study shows that by leveraging transfer learning, a generalized decoding model can be quickly personalized. Specifically, fine-tuning the pre-trained CNN with minimal calibration data from a novel subject produced statistically significant gains in decoding accuracy across all three classification tasks. Importantly, this framework offers practical advantages for clinical use. By reducing per-session calibration to just 10\% of subject data, equivalent to $\sim$3.3 min of EEG recording data, the method alleviates time and fatigue burdens often experienced by subjects with neurological injury. Rapid adaptation makes it feasible to deploy decoders in time-constrained therapy environments or even in home settings.

In Task 1 (EO vs. EC), the high baseline performance reflects well-established spectral differences in the $\alpha$-band power between open- and closed-eye states, echoing prior work on EEG-based cognitive state classification~\cite{barry2007eeg}. When benchmarked against recent transfer-learning work, our approach attained comparable or superior accuracy with a considerably smaller calibration effort. Van der Aar \textit{et al.} required fine-tuning on approximately 40\% of each new night’s data to maximize their sleep-staging CNN's performance \cite{van2024deep}, whereas Wu \textit{et al.} adapted EEGNet~\cite{lawhern2018eegnet} and ShallowCNN~\cite{schirrmeister2017deep} with 2–14\% of each unseen subject's motor-imagery trials reporting peak accuracies of 70–76\% \cite{wu2022transfer}. Using a 10\% calibration subset, our transfer-learned decoder achieved a minimum accuracy of 82.3\%. 

\gc{To test the robustness of our approach, we repeated the above analysis while using the 5\%-95\% and 20\%-80\% update-validate holdout data split. We generally observed similar performance gains as those reported in Table~\ref{tab:accuracy}. The full tabular results are omitted in the interest of brevity. Briefly, for the 5\%-95\% split, the updated models achieved the average performance gains of 5.7, 9.1, and 20.5 percentage points with respect to their baseline counterparts, for Tasks 1, 2, and 3, respectively. Similarly, for the 20\%-80\% split, the corresponding performance gains were 10.8, 17.6, and 25.6 percentage points. While performance gains were observed across all tested proportions of the holdout data, the 10\% and 20\% update conditions yielded comparable accuracy improvements, with diminishing marginal returns beyond 10\%. Notably, even using just 5\% of the holdout data led to substantial improvements over the baseline, highlighting the sensitivity of the proposed update strategy.}

\gc{When compared to the subject-specific SVM classifiers, the updated CNN performances were uniformly superior across tasks (compare Table~\ref{tab:accuracy} and Table~\ref{tab:svm_accuracy}). Specifically, for Tasks 1, 2, and 3, the classification accuracy margins were 4.4, 21.9, and 11.1 percentage points in favor of the updated CNN model, respectively. For completeness, we repeated the 5\%-95\% and 20\%-80\% perturbation analysis with the SVM classifier. At the 5\%-95\% split, the updated CNN outperformed the subject-specific SVM by 8.7, 12.0, and 11.9 percentage points across the three tasks, respectively. At the 20\%-80\% split, the corresponding classification accuracy margins were 3.7, 17.9, and 7.8 percentage points in favor of the updated CNN. These results reinforce the advantage of deep feature learning in more complex classification scenarios such as motor intention decoding and multi-class prediction, where simpler decision boundaries may be insufficient.} 

Together, these findings indicate that spatio-spectral representations learned from previous users can be rapidly and efficiently personalized, strengthening the case for practical plug-and-play BCIs in rehabilitation contexts.

\textbf{Limitations:} Several limitations should be acknowledged. First, the study was conducted exclusively with able-bodied individuals; as such, generalizability to those with disrupted cortical physiology remains unverified. Second, although 10\% of a subject's data was sufficient for fine-tuning in this dataset, the optimal amount of update data was not systematically determined. Third, although the CNN architecture used in this study is lightweight and interpretable, it may not fully capture long-range temporal dependencies or higher-order EEG dynamics that more complex models (e.g., RNNs or Transformers) might exploit. Lastly, the training data were collected under highly controlled, cue-based protocols; real-time, online implementation may involve greater variability and noise.

\textbf{Future Directions:} Future work will focus on expanding the source data pool to encompass a wider range of age and impairments to determine how a more diverse ``super-decoder'' affects transfer learning gains, as performance will likely improve with larger, multi-subject databases~\cite{lotte2018review}. With that backbone in place, further testing can be done to determine the optimal proportion of the update dataset systematically. Additional model architectures will be tested and benchmarked against the current network to determine a suitable architecture for online decoding in clinical settings. Moreover, future work will validate this approach in populations with neurological injury, particularly during clinical trials involving BCI-guided rehabilitation therapy~\cite{biswas2024single}.

\textbf{Conclusion:} In summary, this study provides empirical support for a transfer learning approach that balances generalizability and personalization. With just a small amount of subject-specific data, a pre-trained CNN decoder can be rapidly adapted to new individuals, significantly boosting classification accuracy across cognitive and motor tasks. By minimizing setup time and enhancing flexibility, this strategy has the potential to streamline BCI deployment in clinical environments, supporting broader access to neurotechnology-assisted rehabilitation.

\section*{Acknowledgment}
This study was partially supported by the National Science Foundation (Award \#1646275). 

\bibliographystyle{IEEEtran}
\bibliography{ArXiv}

\clearpage
\end{document}